# The Magnetic Binary GJ 65: A Test of Magnetic Diffusivity Effects

James MacDonald[1], D. J. Mullan[1], and Sergio Dieterich[2]
[1]Dept. Physics & Astronomy, University of Delaware, Newark, DE 19716, U.S.A.
[2]Dept. Terrestrial Magnetism, Carnegie Institution of Washington, Washington, DC 20015, U.S.A

**Abstract**

GJ 65 is a M dwarf binary system consisting of the two flare stars BL Cet (GJ 65A) and UV Cet (GJ 65B). Two teams of investigators have recently reported total magnetic fluxes corresponding to fields of 4.5 and 5.2 kG for GJ65A, and 5.8 and 6.7 kG for GJ65B: for each component, the magnetic results obtained by the two teams agree with each other within $1\sigma$. For the first time, we can directly compare the predictions of our magneto-convective models, based on fitting observed stellar parameters, with measured field strengths. We find that our models agree with the observed field strengths provided the effects of finite conductivity are accounted for. Thus, GJ65 provides us an opportunity to use observations of field strengths to distinguish between the predictions of our models that assume perfect electrical conductivity and those that allow for finite conductivity.

## 1. Introduction

GJ 65 (= L726-8) is a M dwarf binary system consisting of the two flare stars BL Cet (GJ 65A) and UV Cet (GJ 65B) in a wide orbit ($P_{orb}$ = 26.3 – 26.5 yr).

Kochukhov & Lavail (2017) have analyzed the large-scale and local surface magnetic fields of both components of GJ 65. From high-resolution circular polarization spectra, they find the two stars have different global field topologies, which is remarkable because the two stars have essentially the same mass, and very similar radii (Kervella et al. 2016, hereafter K16) and rotation rates (Barnes et al. 2017, hereafter B17). The secondary is found to have an axisymmetric, dipolar-like global field with an average strength of 1.3 kG while the primary has a much weaker, more complex and non-axisymmetric 0.3 kG field. On the other hand, an analysis of the differential Zeeman intensification, which is sensitive to the total magnetic flux, shows the two stars having similar magnetic fluxes of 5.2 and 6.7 kG for GJ65A and B, respectively. These are amongst the strongest surface-averaged field strengths that have been reported for M dwarfs. Based on these complementary magnetic field diagnostic results, Kochukhov & Lavail (2017) suggest that the dissimilar radio and X-ray variability (Audard et al. 2003) of GJ65A and B is linked to their different global magnetic field topologies.

Shulyak et al. (2017) measured total magnetic fluxes corresponding to fields of 4.5 ± 1.0 kG in GJ 65A and 5.8 ± 1.0 kG in GJ 65B. They note that these values are systematically lower than the 5.2 ± 0.5 kG and 6.7 ± 0.6 kG reported by Kochukhov & Lavail (2017) and attribute this to differences in fitting methods. However, the differences in the measurements for each star are not especially significant in a statistical sense: the two values reported for GJ65A differ from each other by an amount that is less than $1\sigma$. The same applies to the two values reported for GJ65B.

K16 have measured by H-band interferometry the angular diameters of the two components of GJ65, $\theta_{UD}$(A) = 0.558 ± 0.008 ± 0.020 mas and $\theta_{UD}$ (B) = 0.539 ± 0.009 ± 0.020 mas. On correcting for limb-darkening, the angular diameters increase by ~3% to $\theta_{LD}$(A) = 0.573 ± 0.021 mas and $\theta_{LD}$(B) = 0.554 ± 0.022 mas. Based on a parallax of 373.70 ± 2.70 mas (van Altena et al. 1995), the linear radii are $R$(A) = 0.165 ± 0.006 $R_\odot$ and $R$(B) = 0.159 ± 0.006 $R_\odot$. From orbital analysis, K16 derived masses for the two components of $M$(A) = 0.1225 ± 0.0043 $M_\odot$ and $M$(B) = 0.1195 ± 0.0043 $M_\odot$, which lead to surface gravities of log $g_A$ (cgs) = 5.092 ± 0.015, and log $g_B$ (cgs) = 5.113 ± 0.015.

K16 note that the radii exceed expectations from stellar structure models by 14 ± 4% and 12 ± 4%, respectively, and propose that this discrepancy is caused by the inhibition of convective energy transport by a strong internal magnetic field generated by dynamo effect in these two fast-rotating stars.

K16 point out that this hypothesis is strengthened by comparing the components of GJ65 with the almost identical but slowly rotating Proxima Cen, which does not appear to be inflated compared to models. From their high-resolution spectra of the two components, K16 determine [Fe/H](A) = -0.03 ± 0.20 and [Fe/H](B) = -0.12 ± 0.20. These values are consistent with the stars having the same heavy element abundance of [Fe/H] ≈ -0.08.

Benedict et al. (2016) have also analyzed the orbit of GJ 65 by combining their HST/FGS measurements of position angle and separation with five measurements with VLT/NaCo (Kervella et al. 2016) and one with HST/NICMOS (Dieterich et al. 2012). They find an orbital period of 26.5 yr that is slightly longer than the 26.3 yr period found by K16. Their mass determinations, M(A) = 0.120 ± 0.003 $M_\odot$ and M(B) = 0.117 ± 0.003 $M_\odot$, are in good agreement with those of K16, and the 0.0025 $M_\odot$ difference in mean values can be attributed to the difference in orbital period.

In a study of surface brightness distributions from Doppler imaging, B17 have measured the stellar radii and find $R_A$ = 0.159 ± 0.010 $R_\odot$ and $R_B$ = 0.160 ± 0.008 $R_\odot$, which are in good agreement with the interferometric radius determinations of K16. It is important to note that B17 also determine that the GJ 65 components are both fast rotators, with rotational periods of 5.83 and 5.44 hrs for A and B, respectively. It seems likely that these fast rotations are in some way responsible for the unusually strong fields that have been reported for both A and B.

In light of these new observations and magnetic field measurements, in this paper we make comparisons with predictions from our model of radius inflation due to magnetic inhibition of convection. We consider models with and without inclusion of finite conductivity effects and find better agreement with the measured field strengths when finite conductivity is taken into account.

A definitive comparison of theory and observation requires accurate determinations of the effective temperatures for each component of the binary, which we provide in section 2. Our method of including magnetic inhibition in stellar structure calculations is described in section 3 and the results of our modelling are given in section 4. In section 5, we discuss how our results depend on adopted values for stellar mass, heavy element abundance, the fraction of the surface that is covered by dark spots, the value of the magnetic field ceiling, and the mixing length ratio. In section 6, we make predictions for the surface magnetic field strengths needed to give the amount of radius inflation that match the observed radii and effective temperature. Our conclusions and discussion are given in section 7.

**2. Luminosity and Temperature Estimates**

There are very few luminosity or temperatures estimates for GJ 65 in the literature. From low-resolution infrared spectra and VRIJHKLL' photometry, Leggett et al. (1996) obtained log $L/L_\odot$ = -2.54 and $T_{eff}$ = 2700 K for the GJ 65AB binary system. In their Fig. 10, K16 compare the positions of GJ65 A and B and Proxima Centauri in the mass-radius diagram to the theoretical isochrones of Baraffe et al. (2015). K16 find that the radii of GJ65 A and B could be consistent with very young stars of age between 200 and 300 Myr. However, K16 also go on to point out that such young ages are incompatible with the observed absolute infrared magnitudes of GJ65 A and B: the infrared brightnesses are too faint compared to the models by approximately 0.3 mag in the JHK bands. In other words, the models predict effective temperatures that are too high by ~7.2%. For ages 200 – 300 Myr, the Baraffe et al. models predict that a 0.12 M star has $T_{eff}$ ~ 2990 – 3020 K which requires that to match the JHK fluxes the actual $T_{eff}$ must be ~ 2790 K. Kochukhov & Lavail (2017) adopted 3000 and 2900 K for GJ65 A and B, respectively. Shulyak et al. (2017) derived $T_{eff}$ ~2900 – 3000 K for component A and ~2800 – 3000 K for component B, using photometric calibrations from Golimowski et al. (2004) and by fitting the strength of the TiO γ-band at 750 nm. However, Shulyak et al. acknowledge that the effective temperatures of M dwarfs are not accurately known, and different sources sometimes list noticeably different values. In summary, $T_{eff}$ values in the literature for the GJ 65 components range from a low of 2700 K to a high of 3000 K.

Colors can also be used to estimate $T_{eff}$. Mann et al. (2015, 2016) give a number of convenient relations that allow estimation of typical M dwarf properties from colors. Using their relation for $T_{eff}$ from V – J and J – H, where J and H are 2MASS magnitudes, we obtain $T_{eff}$ = 2862 ± 27 K. Here the uncertainty is due solely to the uncertainties in the photometry. Mann et al. note that their typical spectroscopic uncertainty is 60 K, and this should be added in quadrature. Mann et al. also provide a relation between $T_{eff}$ and $R$: inserting $T_{eff}$ = 2862 ± 27 K into this relation leads to $R$ = 0.135 ± 0.004 $R_\odot$. This radius estimate is significantly smaller than that measured by K16 or B17. Furthermore, Mann et al. also provide expressions relating $M$ and $R$ to $K_s$: inserting the above radius into the Mann et al. expressions leads to $M$ = 0.149 ± 0.001 $M_\odot$. This mass estimate is significantly larger than the mass determinations of K16. These findings suggest that the components of GJ 65 do not have typical M dwarf properties, which we assume is a consequence of the presence of their intense magnetic fields.

In principle, the effective temperature of a star can also be estimated from its spectral type. From the strengths of TiO bands in the blue, Joy & Abt (1974) assigned GJ 65A (BL Cet) and GJ 65B (UV Cet) to spectral types dM5.5e and dM6e, respectively. From the 7810 Å TiO band, Bessell (1991) assigned GJ 65B spectral type M5.5. From spectra in the range 6500 – 9000 Å, Kirkpatrick, Henry & (1991) determined spectral types M5.5V and M6V for GJ 65 A and B respectively. Using the TiO5 band as their primary spectral-type calibrator, Reid, Hawley & Gizis (1995) determined spectral type M5.5 for GJ65A. From the K-band spectrum, Boeshaar & Davidge (1994) determined spectral type dM6- for the GJ 65 binary. Hence, the spectral types of the two stars are well constrained by both visible and IR spectra to be within the range from M5.5 to M6.

However, for specific stars of known spectral type, it is noteworthy that there often is a large range of $T_{eff}$ estimates in the literature. For example, K16 compare GJ 65A and B to Proxima Centauri, which may (according to some reports) be of similar spectral type: M5.5Ve (Bessell 1991), M5.5 (Henry et al. 2002), M6Ve (Torres et al. 2006). On the other hand, dissimilar spectral types have also been reported: e.g. M5 (Lurie et al. 2014), and M7 (Gaidos et al. 2014). A large number of estimates of $T_{eff}$ for Proxima appear in the literature, ranging from a lowest value of 2425 K (Leger et al. 2015) to a highest value of 3083 K (Loyd & France 2014). The range in these $T_{eff}$ values (spanning almost 660 K) indicates that the task of obtaining $T_{eff}$ from spectral type is subject to large uncertainty. Similar large ranges in $T_{eff}$ determinations are found for other stars of spectral type M5.5 and M6. For example, the close binary GJ 1245AC, which has been used as a standard for spectral type M5.5, has $T_{eff}$ determinations ranging from a lowest value of 2661 K (Huber et al. 2014) to a highest value of 4064 K (Wright et al. 2011). As a second example, for the M6 standard Wolf 359, the $T_{eff}$ determinations from the literature range from 2500 K (Casagrande, Flynn, & Bessell 2008) to 3356 K (Terrien et al. 2015). Averaging the $T_{eff}$ values from the literature, we obtain $T_{eff}$ = 2843 ± 203 K for a star that is regarded as a "standard" for stars of spectral type M6.

Recently, three studies have addressed the issue of M dwarf effective temperatures based on comparing data to synthetic spectra produced by model atmospheres (Rajpurohit et al. 2013; Dieterich et al. 2014; Mann et al. 2015). All three use the latest generation BT-Settl model atmospheres of Allard et al. (2012b, 2013), which account for particulate sedimentation using an adaptive cloud model and use the Caffau et al. (2011) revised solar abundances.

Rajpurohit et al. (2013) take the most direct approach by comparing the synthetic spectra to observed spectra from spectral standards through least squares minimization. They apply the technique to spectra taken with ESO Multi Mode Instrument on the 3.6-m New Technology Telescope (NTT) at La Silla, with a wavelength range of 5200 Å to 9500 Å and also with the Double Beam Spectrograph at Siding Spring Observatory with a wavelength range of 3000 Å to 10000 Å. Their Figures 2 and 3 show excellent overall agreement between model spectra and observed spectra throughout the M dwarf sequence. However, they note that missing opacities due to several molecular species cause discrepancies in the bluer part of the model spectra (Rajpurohit et al. 2013, Section 5.1). Rajpurohit et al. (2013) assign effective temperatures based on the best fitting grid element, without interpolation, and assign an uncertainty of 100 K to account for the grid spacing. Their Table 1 lists temperatures of 2900 K for three M5 dwarfs, 2800 K for four M5.5 dwarfs, 2800 K for two M6 dwarfs, and 2700 K for four M6.5 dwarfs as well as three M7 dwarfs.

Dieterich et al. (2014) calculated effective temperatures for cool M dwarfs and L dwarfs by comparing the same BT-Settl models to photometric data spread over a broad range of wavelengths. For each object, they obtain a total of 21 different colors consisting of all combinations for which the bluer band is V, R, or I by combining their own VRI photometry on the Bessel system with 2MASS JHKs magnitudes (Skrutskie et al. 2006) and WISE W1, W2, and W3 photometry (Wright et al. 2010). These colors cover the spectral range from ∼0.4 μm to ∼16.7 μm. The same 21 colors are calculated for each spectrum in the BT- Settl model atmosphere grid (Allard et al. 2012b, 2013). The best matching $T_{eff}$ is found by interpolating $T_{eff}$ as a function of the residuals of the observed color − synthetic color comparison to the point of zero residual. Since the technique can be applied independently to each available photometric color, the adopted $T_{eff}$ for an object is the mean of the $T_{eff}$ values from each color. A measure of the uncertainty in $T_{eff}$ is obtained from the standard deviation of the ensemble of $T_{eff}$ values. This technique has the advantage of sampling ~97% of the SED from the visible to the mid infrared, and of providing multiple independent calculations of $T_{eff}$ values. Further, because any imperfections in the models such as missing opacities are not likely to affect the entire SED range the dispersion around the true $T_{eff}$ value more closely approximates a random distribution that can be used as an estimate of the uncertainty. The mean $T_{eff}$ of the six M6 dwarfs listed in Dieterich et al. (2014) is 2713 K with a standard deviation of 109 K. These values are a measure of the intrinsic dispersion in temperatures to be expected in any M subtype. The mean of the individual uncertainties is 22 K and is a measure of the method's precision. Table 6 of Dieterich et al. (2014) lists objects in common with Rajpurohit et al. (2013) and shows that in general for late M dwarfs the temperatures of Dieterich et al. (2014) are about 100 K cooler than those of Rajpurohit et al. (2013). As a check on our methodology we also calculated the effective temperature and luminosity of GJ 65 using a modified version of the Dieterich et al. (2014) methodology and obtained $T_{eff}$ = 2808 K ± 33 K, and L/L$_\odot$ = 0.00281 ± 0.00007. The best fit was obtained for the log $g$ = 4.5 atmosphere models. However, at these temperatures, the color fits depend very weakly on log $g$, and are also consistent with log $g$ = 5. In this updated methodology, we use only the colors that show a good match between synthetic model photometry and the observed sequence of M and L dwarfs in color-color plots. In order to be selected as a good color for $T_{eff}$ estimation the color in question must be present in at least two color-color diagrams that reproduce the observed color sequence and the other color in the color-color diagram must not have bands in common with the color being tested. Details of the color vetting process will be discussed in an upcoming paper[1]. Figure 1 shows how the color residuals converge to zero at this temperature.

Finally, Mann et al. (2015) calculated effective temperatures for a large sample of main sequence stars ranging in spectral type from K7 to M7. Their mean temperature for the six objects ranging in spectral type from M5.5 to M6.5 in their spectral type scale, which is binned to the nearest tenth, is 2866 K with a standard deviation of 47 K. Their standard uncertainty for individual objects is 60 K. Their methodology is described in detail in Mann et al. (2013). The sample includes the well-known M6 red dwarf Wolf 359 (GJ 406) for which they assign $T_{eff}$ = 2818 ± 60 K. The other temperatures for objects in the M6 range are: 2864 K (PM I00115+5908), 2829 K (GJ 3146), 2880 K (GJ 3147), 2951 K (PM I10430-0912), and 2859 K (GJ 1245 B). These temperatures are systematically higher than those of

---

[1] The colors found to represent the observed sequence well are: V - R, V - H, V - K, V - W1, V - W2, R - H, R - K, R - W1, I - K, and I - W1.

Rajpurohit et al. (2013) and especially those of Dieterich et al. (2014). The higher temperatures for mid M dwarfs in Mann et al. (2015) are probably a bias effect for two reasons. First, their method uses linear interpolation of the three best model spectra to calculate properties but their model grid has $T_{eff}$ increments of 100 K with the lowest temperature being 2700 K. Any result in the 2700 K to 2800 K temperature range may therefore be biased by the lack of a suitable lower temperature model spectrum in the interpolation. Second, their comparisons to the BT-Settl models are based only on red optical spectra ranging from 5500 Å to 9300 Å and employ a weighing scheme that discards several wavelength regions where mean discrepancies for the entire sample are greater than 10 percent (Mann et al. 2013, Figure 9). Discarding small regions of discrepancy is justifiable in principle, but it must be done as a function of temperature and not as a single fit for a sample that ranges in spectral type from K7 (4100 K) to M7 (2700 K). The relative strengths of several opacities change significantly in this temperature range and the selected wavelength windows may not be a good match to temperatures close to the lower bound of the range. For these reasons, we believe that the effective temperatures of Rajpurohit et al. (2013) and Dieterich et al. (2014) are more realistic for M6 dwarfs.

If we assume that GJ 65A and B have the same effective temperature and luminosity, then the above $T_{eff}$ and $L$ for the system results in a component radius $R = 0.159$ R$_\odot$ that is in excellent agreement with the radius measurements of B16 and K17.

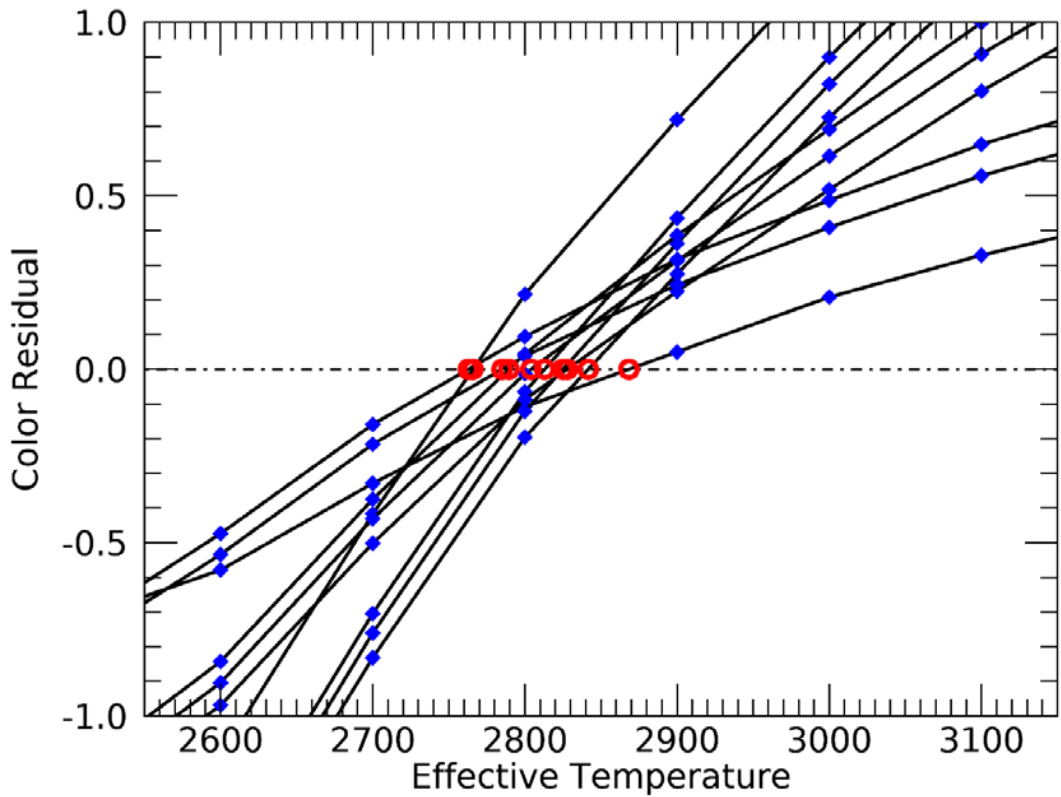

Figure 1. Effective temperature interpolations as a function of color for GJ 65. Each black line represents one of ten photometric colors. Blue diamonds denote the residuals when comparing the model spectrum at the indicated temperature to the observed color. The open red circles denote the interpolation at the point in which the model color matches the observed color. We derive $T_{eff}$ = 2808 K ± 33 K based on the mean and standard deviation of the interpolated color matches.

In our analysis of the GJ 65 binary system so far, we have used the unresolved 2MASS JHKs photometry (Skrutskie et al. 2006). K16 note that their combined magnitudes in the JHK bands are systematically fainter by ~ 0.15 mag than the 2MASS values. K16 suggest this cause of this difference might be due to long- term variability of the two stars. When the 2MASS observations were made in 1998, the two stars were near periapsis and their separation (~ 1") was significantly less than ~ 2" separation when the K16 observations were made. Melikian et al. (2011) have found that the flare rate in GJ 65 was greatest at the time of the previous periapsis (late 1971 or early 1972). K16 suggest that the two stars were more active and brighter in the infrared at the time of the 2MASS observations. Independent reports have also suggested that the brightness of UV Cet (or more strictly GJ65) has varied from one observing season to the next by amounts ranging from 0.1 to 0.3 mag (Migach 1965; Chuadze 1965; Karamish et al. 1966). In view of such reports, a reliable $T_{eff}$ estimate based on the K16 radii should be based on contemporaneous photometry. A reduction in the stellar flux by 0.15 mags corresponds to a reduction in $T_{eff}$ of ~ 3.4%, and we expect $T_{eff}$ based on the NaCo observations to be ~ 2700 K.

The discrepancy in JHK magnitudes could also be due to the fact that the 2MASS point source photometry algorithm assumes a single, non-blended point source when using a curve of growth correction for aperture photometry (2MASS Explanatory Supplement https://www.ipac.caltech.edu/2mass/releases/allsky/doc/sec4_4c.html). A separation of ~1" may not be enough so that two distinct sources are identified, but may be wide enough to cause erroneous PSF centroiding and application of the curve of growth.

In order to determine the individual stellar temperatures and luminosities from published photometry, we use a multi-step process. The distance is determined by combining the parallax measurements of van Altena et al. (1995) and Dupuy & Liu (2012) to get $\pi$ = 0.3741 ± 0.0022" and distance 2.673 ± 0.016 pc. We first estimate the composite luminosity of the binary system by using fluxes from the BVRI magnitudes from Winters et al. (2015), and the WISE magnitudes (Dupuy & Liu 2012) for the system, and the sum of the fluxes from the individual star $JHK_sL'$ magnitudes from K16, and the HST WFPC2 1042 nm magnitudes (Schroeder et al. 2000). We convert the apparent magnitudes to fluxes by using the zero magnitude fluxes from the Spanish Virtual Observatory Filter Profile Service[2]. Our results are given in Table 1.

Table 1. Calculated fluxes for the (unresolved) composite GJ 65AB binary system

| Band | Wavelength (µm) | Apparent magnitude | log $f$ (erg s$^{-1}$ cm$^{-2}$ µm$^{-1}$) |
|---|---|---|---|
| B | 0.4378 | 12.78 ± 0.05 | -9.313 ± 0.020 |
| V | 0.5466 | 12.06 ± 0.02 | -9.279 ± 0.012 |
| R | 0.6696 | 10.40 ± 0.05 | -9.439 ± 0.020 |
| I | 0.7862 | 8.34 ± 0.05 | -8.499 ± 0.020 |
| HST/WFPC2.f1042m | 1.0435 | 6.91 ± 0.07 | -8.018 ± 0.028 |
| J | 1.24775 | 6.42 ± 0.07 | -8.099 ± 0.028 |
| H | 1.63725 | 5.84 ± 0.05 | -8.279 ± 0.020 |
| K | 2.14206 | 5.52 ± 0.07 | -8.564 ± 0.028 |
| W1 | 3.3526 | 5.05 ± 0.07 | -9.103 ± 0.028 |
| L' | 3.76978 | 5.17 ± 0.075 | -9.358 ± 0.030 |
| W2 | 4.6028 | 4.57 ± 0.04 | -9.442 ± 0.010 |
| W3 | 11.5608 | 4.76 ± 0.02 | -11.052 ± 0.005 |
| W4 | 22.0883 | 4.62 ± 0.03 | -12.135 ± 0.0075 |

---

[2] The SVO Filter Profile Service. Rodrigo, C., Solano, E., Bayo, A. http://ivoa.net/documents/Notes/SVOFPS/index.html, The Filter Profile Service Access Protocol. Rodrigo, C., Solano, E. http://ivoa.net/documents/Notes/SVOFPSDAL/index.html

We integrate the fluxes from B to W4 by using monotonic spline interpolation (Steffen 1990) to find flux as a function of wavelength. Using the distance from the parallax, we find $L_{AB} = 0.00272 \pm 0.00007$ $L_\odot$. If we assume, based on their similar spectral types, that the two components have the same temperature, this luminosity coupled with the K16 radius measurements gives effective temperature, $T_{eff} = 2757 \pm 41$ K. As a check on this method, we have also performed the same calculation but with the 2MASS JHK fluxes instead of the NaCo fluxes. We find $L_{AB} = 0.00300 \pm 0.00006$ $L_\odot$ and $T_{eff} = 2824 \pm 39$ K. Our simple method of integrating the fluxes gives a $T_{eff}$ value only 0.6% higher than that found by the more sophisticated method of Dieterich et al. (2014). This gives us confidence in the reliability of our method.

Our next step is to estimate the individual luminosity of each component. For each star, there are results for the individual components in data obtained by NaCo in the filters $JHK_sL'$ and in data obtained by HST WFPC2 in the 1042 nm filter: these filters span the wavelength range 1.0435 to 3.770 µm. The fluxes are given in Table 2. For the J, H and $K_s$ bands, the difference in apparent magnitudes is the same, $m_A - m_B = -0.16$ but for the f1042m band $m_A - m_B = -0.21$ and for the L' band $m_A - m_B = -0.095$, which suggests that GJ 65B is slightly cooler than GJ 65A, consistent with the spectral type of GJ65A being somewhat earlier than that of GJ65B.

Table **2**. Calculated fluxes for GJ 65A and B

| Band | Wavelength (µm) | Apparent magnitude | | $\log f$ (erg s$^{-1}$ cm$^{-2}$ µm$^{-1}$) | |
| --- | --- | --- | --- | --- | --- |
| | | A | B | A | B |
| f1042m | 1.0435 | 7.56 ± 0.05 | 7.77 ± 0.05 | -8.280 ± 0.020 | -8.365 ± 0.020 |
| J | 1.24775 | 7.10 ± 0.05 | 7.26 ± 0.05 | -8.369 ± 0.020 | -8.433 ± 0.020 |
| H | 1.63725 | 6.515 ± 0.05 | 6.675 ± 0.05 | -8.549 ± 0.014 | -8.613 ± 0.014 |
| K | 2.14206 | 6.20 ± 0.05 | 6.36 ± 0.05 | -8.834 ± 0.020 | -8.898 ± 0.020 |
| L' | 3.76978 | 5.885 ± 0.075 | 5.980 ± 0.075 | -9.642 ± 0.021 | -9.680 ± 0.021 |

By integrating the fluxes from 1.0435 to 3.770 µm, we obtain an integrated flux ratio $L_B / L_A$ $0.864 \pm 0.028$. Combining this with our composite luminosity estimate for the system gives $L_A = 0.00146 \pm 0.00004$ $L_\odot$, and $L_B = 0.00126 \pm 0.00004$ $L_\odot$. The corresponding temperatures are $T_{eff,A} = 2781 \pm 55$ K, $T_{eff,B} = 2731 \pm 56$ K when the K16 radii are used and $T_{eff,A} = 2835 \pm 92$ K, $T_{eff,B} = 2725 \pm 72$ K when the B17 radii are used. Using these temperatures as a guide, we next estimate how much of the total flux lies outside the wavelength range 1.0435 to 3.770 µm, by using the BT_Settl model spectra (Allard et al. 2012a) for $T_{eff} = 2700$ and 2800 K, $\log g = 5$. We find that 25.4 and 27.5% of the flux lies outside this range for $T_{eff} = 2700$ and 2800 K, respectively. Adding a temperature dependent correction to include this flux, we obtain $L_A = 0.00147 \pm 0.00005$ $L_\odot$, and $L_B = 0.00125 \pm 0.00005$ $L_\odot$. The corresponding temperatures are $T_{eff,A} = 2784 \pm 58$ K, $T_{eff,B} = 2728 \pm 60$ K when the K16 radii are used and $T_{eff,A} = 2842 \pm 98$ K, $T_{eff,B} = 2715 \pm 81$ K when the B17 radii are used.

## 3. Modelling Technique

Our modelling technique has been recently described in MacDonald & Mullan (2017b). We first construct stellar models of appropriate mass and age that do not include any effects due the presence of magnetic fields. Because many authors, including K16, compare their results to the M dwarf models of Baraffe et al. (2015), in the present paper, we use the BT-Settl atmosphere models[3] (Allard et al. 2012a, 2012b; Rajpurohit et al. 2013) in order to provide the outer boundary conditions for this set of models.

---

[3] https://phoenix.ens-lyon.fr/Grids/BT-Settl/CIFIST2011/

Because the BT-Settl atmospheres do not include any magnetic effects, and because magnetic effects are the primary motivation of our modeling approach, we introduce a second step in the process: we construct non-magnetic models using boundary conditions from atmosphere calculations (for details see MacDonald & Mullan 2017b) based on the $T - \tau$ relation of Krishna-Swamy (1966) that match the photospheric properties of the models based on the BT-Settl atmosphere boundary conditions. This step requires us to adjust the mixing length parameter, $\alpha$. The value $\alpha$ of that gives the best match to the BT-Settl boundary condition models is found to vary with $\log g$ and $T_{eff}$. For simplicity, we use a single value for $\alpha$ appropriate for low mass main sequence models, $\alpha = 0.7$. The impact on our results from adopting a different value for $\alpha$ is discussed in section 5. Finally, as a third step in the process, we construct models that include magnetic effects consistently in the Krishna-Swamy atmosphere and in the stellar interior. Because of the low effective temperatures of the GJ 65 components, we calculate models that include the effects of finite magnetic diffusivity effects in addition to models that assume zero magnetic diffusivity. We refer to the models with zero diffusivity as GT models and the models with finite diffusivity as GTC models (for details, see Mullan & MacDonald 2010, MacDonald & Mullan 2017b).

We stress that we use the Krishna-Swamy atmospheres only to allow us to include magnetic inhibition of convection, a process that is not included in the BT_Settl atmospheres. We do not use Krishna-Swamy atmospheres for determining SEDs, and comparison with observations.

Our adopted magnetic field profile depends on two-parameters: A magnetic inhibition parameter, $\delta$ and a magnetic field ceiling $B_{ceil}$. The magnetic inhibition parameter, which was first introduced by Gough & Tayler (1966), is a local parameter defined by

$$\delta = \frac{B_v^2}{4\pi\gamma P_{gas} + B_v^2}, \qquad (1)$$

where $B_v$ is the vertical component of the magnetic field and $P_{gas}$ is the gas pressure. In the original Gough & Tayler criterion, convective stability is ensured as long as the radiative gradient $\nabla_{rad}$ does not exceed $\nabla_{ad} + \delta,$ where $\nabla_{ad}$ is the adiabatic gradient. We have modified the Gough & Tayler criterion to allow for deviations from an ideal gas and to include the effects of finite electrical conductivity.

Our approach in computing magneto-convective models is to assume that in any location within a stellar convection zone, the local field is aligned with the only preferred direction in the problem, i.e. that defined by the gravity vector. (This essentially 1-D approach is standard in any stellar model that relies on mixing length theory.) Therefore, in a strict sense, we assume that the dynamical effects generated by the presence of a magnetic field are governed by the component of the field that is aligned with the gravity vector, i.e. the vertical component of the field. Of course, in an actual star, the field is 3-D, and will in general contain poloidal and/or toroidal components. Therefore, in any 1-D modeling effort that includes magnetic fields, the numerical values of $B_V$ and $B_{ceil}$ in our code correspond to the unsigned spherically averaged vertical magnetic field strength.

In our models, $\delta$ is kept fixed until a depth is reached at which the vertical field strength increases to a value that is equal to the value we have chosen for the ceiling strength: $B_v = B_{ceil}$. At deeper depths, $\delta$ is determined by the condition that $B_v$ is kept equal to $B_{ceil}$. For the models used here, we choose $B_{ceil} = 10$ kG. This choice of ceiling on the magnetic field strength is motivated by simulations of dynamos in low mass stars. Browning (2008) found that turbulent convective motions in a model that rotates with the same period as the Sun (~27 days) can generate fields with a maximum strength of 13.1 kG (listed in Browning's Table 2). For a model of a completely convective star with a rotation period of 20 days,

Yadav et al. (2015) found the maximum field strength is 14 kG. In the case of GJ65AB, the rotation periods (0.24 and 0.23 days: B17) are 2 orders of magnitude faster than either Browning (2008) or Yadav et al. (2015) have considered. In principle, the fields generated by dynamo action in GJ65AB could very well exceed the values of 13.1 - 14 kG reported by the above authors. However, in the absence of turbulent convective models at periods as short as 0.2 days, we adopt a conservative approach and assume that values of 10 - 20 kG are plausible values for $B_{ceil}$. If subsequent evidence becomes available to suggest that larger $B_{ceil}$ values are appropriate in GJ65AB, the values we quote below for surface field strengths will need to be reduced by a factor proportional to $B_{ceil}^{-0.07}$ (MacDonald & Mullan 2017a).

## 4. Model Results

In figure 2, we show the evolutionary tracks in the $T_{eff} - R$ plane for [Fe/H] = 0.0, $M$ = 0.12 M$_\odot$ models. The tracks end at an age equal to that of the Universe, i.e. 13.7 Gyr.

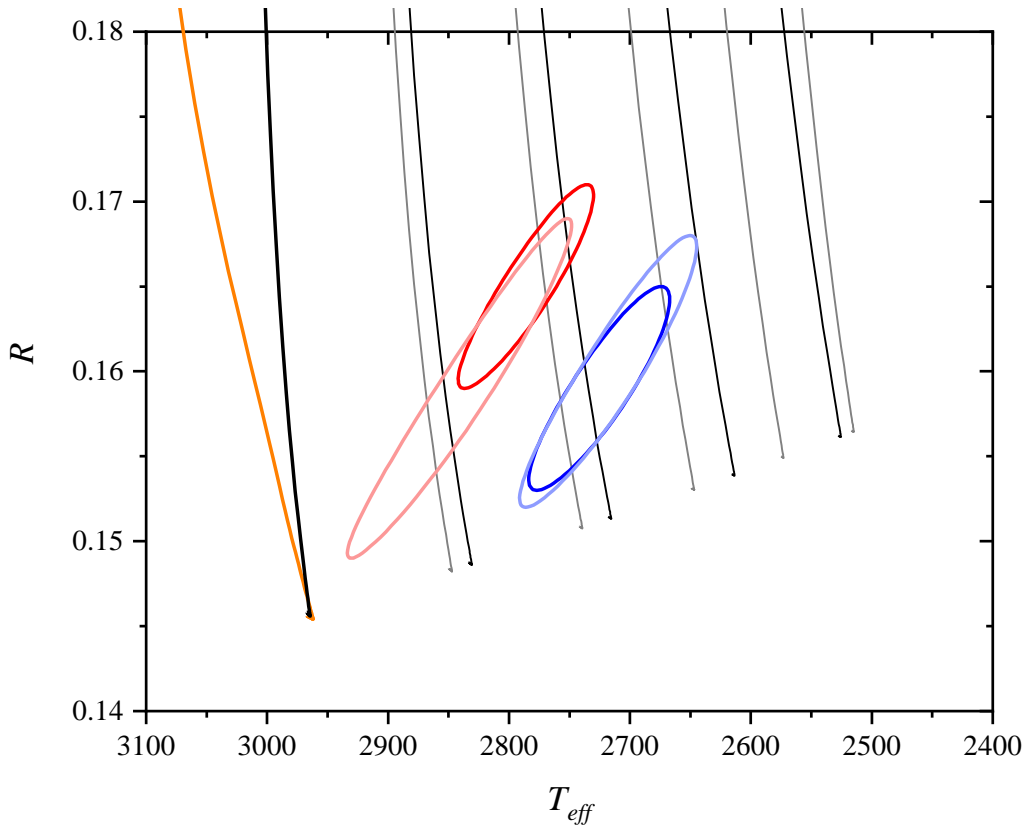

Figure 2. Evolutionary tracks in the $T_{eff} - R$ plane for [Fe/H] = 0.0, $M$ = 0.12 M$_\odot$ models. The orange line is the track for non-magnetic models calculated with the BT_Settl outer boundary condition. The black lines are our GT models tracks for $\delta$ values ranging from 0.0 (leftmost) to 0.08 (rightmost) in increments of 0.02. The grey lines are our GTC models tracks for $\delta$ values ranging from 0.0 to 0.20 in increments of 0.04. The red and blue ovals show the 1$\sigma$ bounds on the locations of GJ 65A (BL Cet) and GJ 65B (UV Cet) respectively, with the darker colors corresponding to the K16 radii measurements and the lighter colors to those of B17.

If it were the case that no estimates of $T_{eff}$ were available for GJ65AB, then the inflated radii of both components could be attributed solely to 'youth'. In such a case, the K16 radius measurements alone would be consistent with ages 170 – 280 Myr for GJ 65A and 180 – 350 Myr for GJ 65B, and, in the same case, for the B17 radius measurements, we would obtain ages > 180 Myr for GJ 65A and 170 – 370 Myr for GJ 65B.

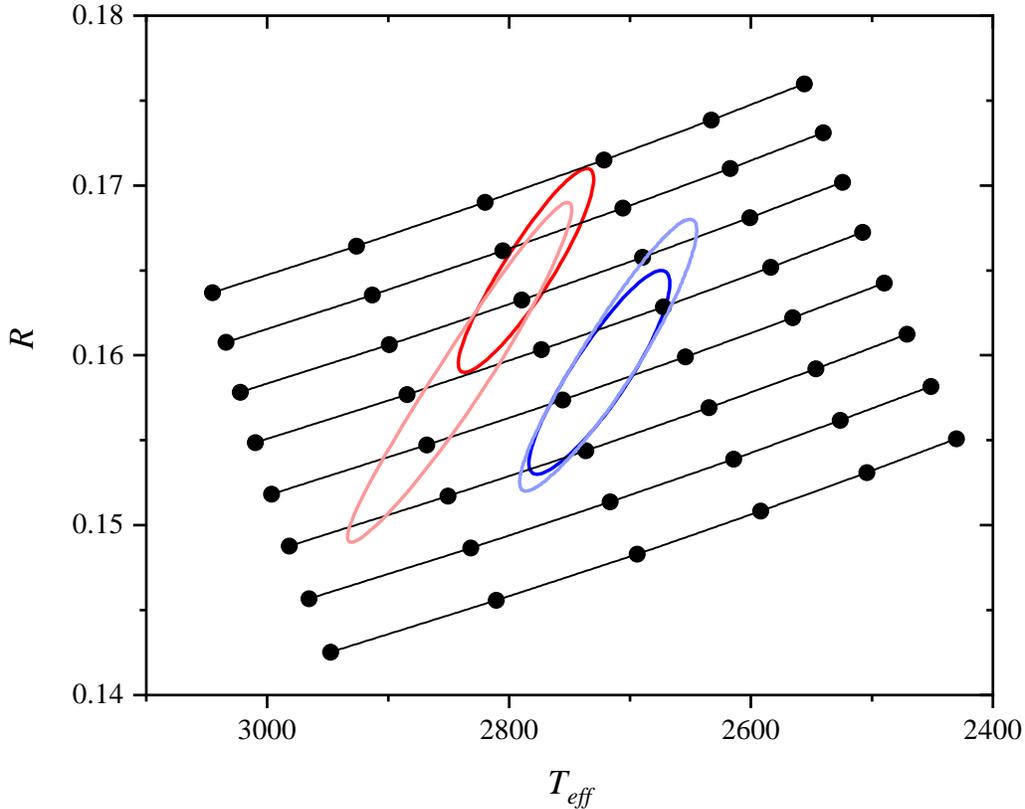

Figure 3. Results of magneto-convective models based on the GT criterion (i.e. zero magnetic diffusivity). Model $T_{eff} - R$ isochrones are compared with the location of the GJ 65 components based on the K16 and B17 radius measurements. The lines show the 5 Gyr [Fe/H] = 0.0 isochrones for $M$ = 0.117 to 0.138 $M_\odot$ in increments of 0.003 $M_\odot$. The filled circles indicate the value of $\delta$, which increases from left to right from 0.00 to 0.10 in increments of 0.02.

However, as we have described in Section 2, information on the $T_{eff}$ values is available for both components of GJ65AB. The empirical values that we have obtained for $T_{eff}$ (see ovals in Figure 2) point clearly to the fact that non-magnetic models (see the left-most orange and black lines in Figure 2) are not consistent with our empirical $T_{eff}$ values for either GJ65A or GJ65B. Moreover, the presence of the strong magnetic fields measured by Kochukhov & Lavail (2017) and Shulyak et al. (2017) implies that the radii are likely inflated due to magnetic inhibition of convection and/or by the presence of surface dark spots. These magnetic effects have the additional consequence of reducing $T_{eff}$, compared to the $T_{eff}$ value of a non-magnetic star of equal mass and age. Therefore, the availability of a $T_{eff}$ measurement contributes significantly to an ability to distinguish between the effects of youth alone and the effects due to the presence of a magnetic field.

    In figure 3, we compare main sequence $T_{eff} - R$ isochrones with the location of the GJ 65 components based on the K16 and B17 radius measurements. We take an age of 5 Gyr as representative of the main sequence. Our results are not sensitive to the adopted age because of the long main sequence lifetimes (~$3\times10^4$ Gyr) for stars with the masses considered here. The solid lines show the 5 Gyr [Fe/H] = 0.0 GT isochrones for $M$ = 0.117 to 0.138 $M_\odot$ in increments of 0.003 $M_\odot$. (We have chosen this range of

model masses to ensure that we cover most of the empirical masses: according to K16, the 1$\sigma$ upper limit on the mass of GJ65A is 0.127 M$_\odot$, while the 1$\sigma$ lower limit for GJ65B is 0.115 M$_\odot$.) The filled circles indicate the value of $\delta$, which increases from left to right from 0.00 to 0.10 in increments of 0.02. Based on the 1$\sigma$ limits on radius, we see that for GJ65A to be on the main sequence requires its mass to be between 0.129 and 0.138 M$_\odot$ for the K16 radius, and between 0.122 and 0.136 M$_\odot$ for the B17 radius. Hence the K16 mass and radius measurements are only marginally consistent with GJ65A being a main sequence star. The B17 radius measurements for GJ 65A are consistent with both the K16 and Benedict et al. (2016) mass determinations. The K16 and B17 radius measurements for GJ 65B are consistent with main sequence models for masses in the ranges 0.123 to 0.131 M$_\odot$ and 0.122 to 0.133 M$_\odot$, respectively. These mass ranges are consistent at the 1$\sigma$ level with the K16 mass determination but not with the Benedict et al. (2016) value. Inspection of Fig. 2 shows that in order to replicate the K16 empirical $R$ and $T_{eff}$ values by means of GT magnetic models, the value of $\delta$ must be chosen to lie in the range 0.03 - 0.06. These values of $\delta$ will enable us to estimate the surface field strengths (see Table 5 below).

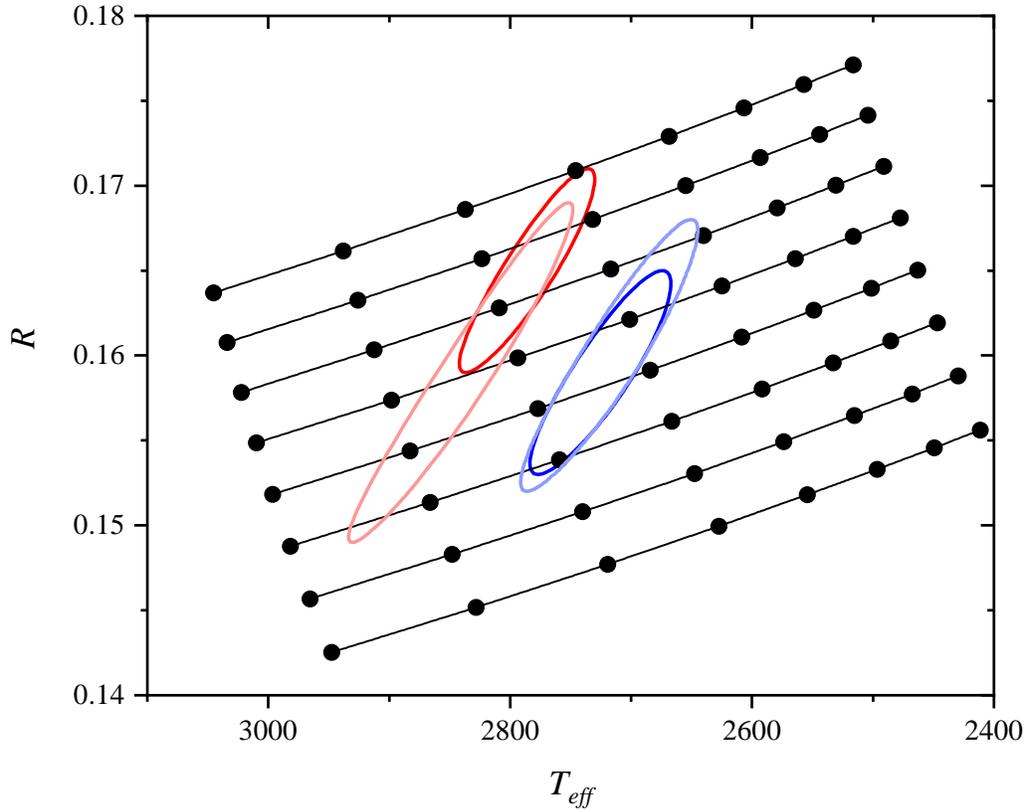

Figure 4, as figure 3 except that the magneto-convective models are based on the GTC criterion (i.e. finite magnetic diffusivity). The filled circles indicate the value of $\delta$, which here increases from left to right from 0.00 to 0.28 in increments of 0.04.

Figure 4 is the same as figure 3 except our GTC models are used. Here the filled circles indicate the value of $\delta$, which increases from left to right from 0.00 to 0.28 in increments of 0.04. We see that the main difference between our GT and GTC models is that, as a consequence of finite magnetic diffusivity (which allows the field lines to move relative to the matter), the GTC models require a larger $\delta$ value to achieve the same degree of radius inflation as the GT models. Specifically, inspection of Fig. 4 shows that in order to have the GTC models replicate the K16 empirical values, the value of $\delta$ must be chosen to be in the range 0.06 - 0.14. These larger values of $\delta$ correspond to stronger fields on the surface of the stars when finite magnetic diffusivity is included (see Table 5).

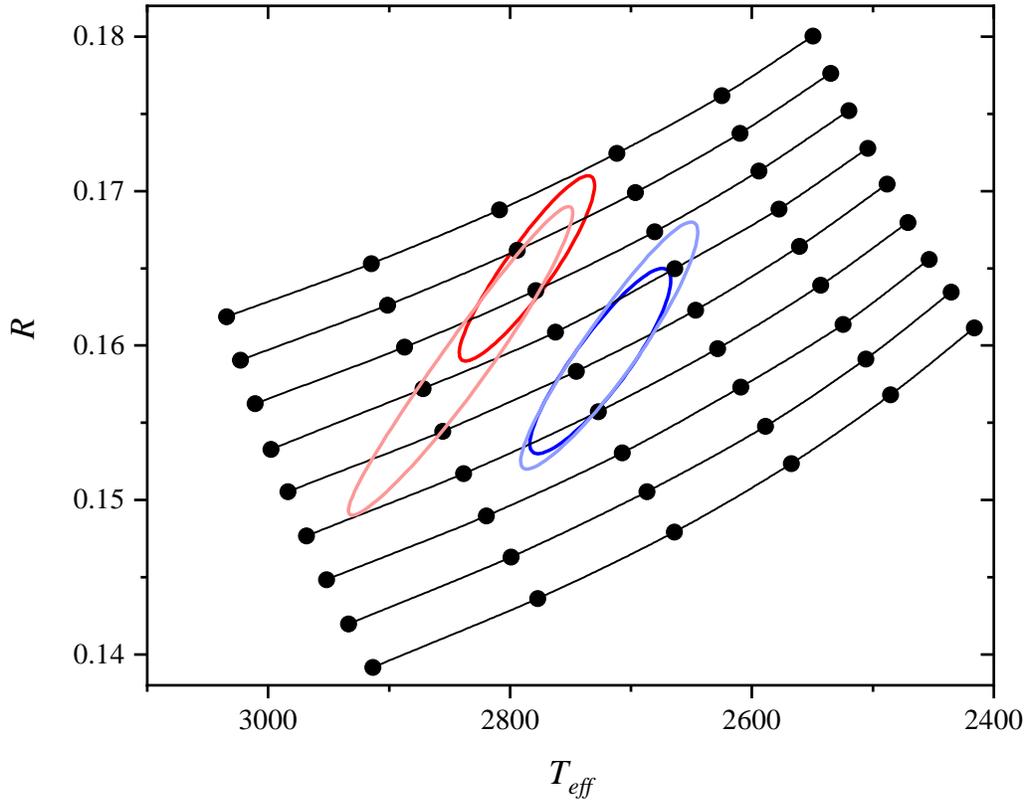

Figure 5. Pre - main sequence GT model $T_{eff} - R$ isochrones compared with the location of the GJ 65 components based on the K16 and B17 radius measurements. The lines show the 500 Myr [Fe/H] = 0.0 isochrones for $M = 0.111$ to 0.135 $M_\odot$ in increments of 0.003 $M_\odot$. The filled circles indicate the value of $\delta$, which increases from left to right from 0.00 to 0.10 in increments of 0.02.

Turning our attention now to pre-main sequence magneto-convective models, in figures 5 and 6, we compare pre-main sequence $T_{eff} - R$ isochrones with the location of the GJ 65 components. In figure 5, the solid lines show the 500 Myr [Fe/H] = 0.0 GT isochrones for $M = 0.111$ to 0.135 $M_\odot$ in increments of 0.003 $M_\odot$. The filled circles indicate the value of $\delta$, which increases from left to right from 0.00 to 0.10 in increments of 0.02. Based on the $1\sigma$ limits on radius, we see that for GJ65A to be a pre-main star of age 500 Myr requires its mass to be between 0.126 and 0.134 $M_\odot$ for the K16 radius, and between 0.120 and 0.133 $M_\odot$ for the B17 radius. Hence all the mass and radius measurements are consistent with GJ65A being a pre - main sequence star. The K16 and B17 radius measurements for GJ 65B are consistent with pre - main sequence models for masses in the ranges 0.119 to 0.126 $M_\odot$ and 0.118 to 0.128 $M_\odot$, respectively. These mass ranges are consistent at the $1\sigma$ level with both the K16 and Benedict et al. (2016) mass determinations.

Is there any independent evidence that GJ 65 is in the pre-main sequence phase? Based on space velocity measurements, Montes et al. (2001) list GJ 65B as a possible member of the Hyades supercluster, which based on the Hyades cluster has an age of $625 \pm 50$ Myr (Perryman et al. 1998).

Comparison of Fig. 5 with Fig. 3, and comparison of Fig. 6 with Fig. 4, indicates that, whether we consider pre-main sequence ages or main sequence ages for the stars in GJ65, essentially the same ranges of $\delta$ values are required for magnetic models (whether for GT or for GTC) to fit the empirical $R$ and $T_{eff}$ values. Specifically, GT models require $\delta \approx 0.01 - 0.07$, while GTC models require $\delta \approx 0.02 - 0.16$.

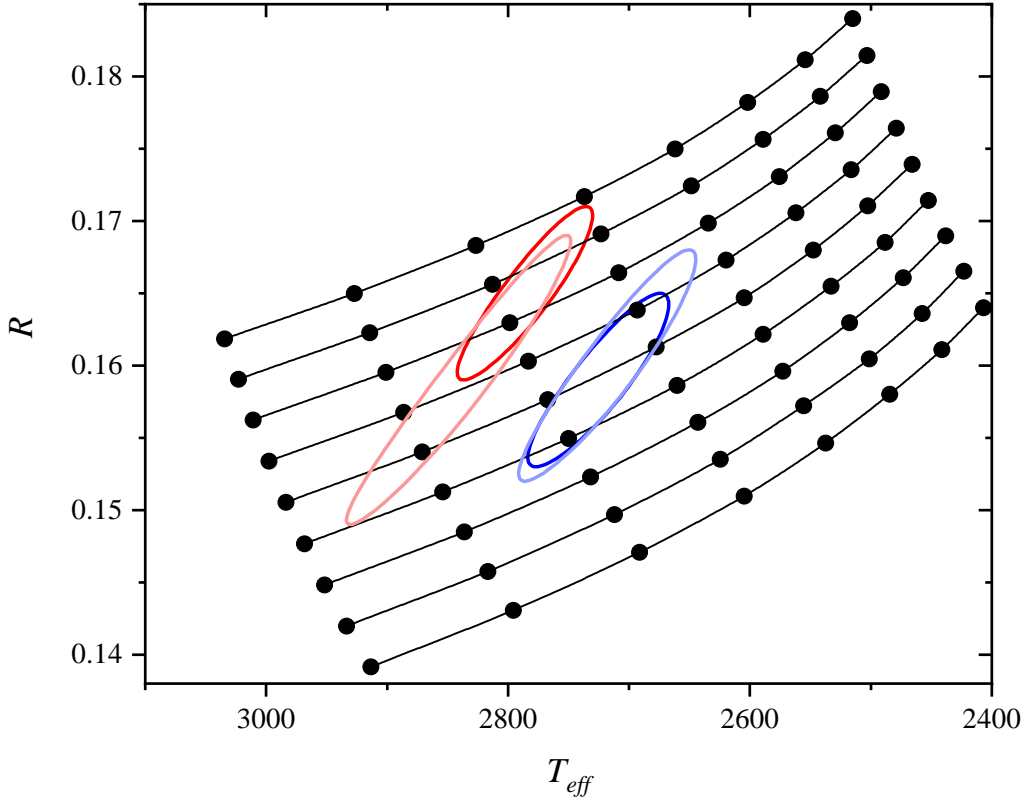

Figure 6, as figure 5 except for magneto-convective models based on the GTC criterion. The filled circles indicate the value of $\delta$, which here increases from left to right from 0.00 to 0.28 in increments of 0.04.

## 5. Dependence on Mass, Heavy Element Abundance, Spots, Magnetic Field Ceiling, and Mixing Length Ratio

The evolutionary tracks shown in figure 2 have been computed for a particular mass, $M = 0.12$ $M_\odot$ and a particular heavy element abundance, [Fe/H] = 0.0. In addition, we have assumed a surface spot coverage fraction, $f_s = 0$, and magnetic field ceiling of $10^4$ G. The mixing length ratio for our models with Krishna – Swamy atmosphere outer boundary conditions has been chosen to match models with the BT_Settl atmosphere outer boundary condition. In this section, we consider how variations in these parameters affect our results.

The primary GJ 65A is 2.5% more massive than the secondary GJ 65B. For main sequence stars of $M \sim 0.12$ $M_\odot$, our models predict this mass difference gives rise to a radius difference of 2.2%.

K16 determined heavy element abundances [Fe/H](A) = -0.03 ± 0.20 and [Fe/H](B) = -0.12 ± 0.20. If we make the reasonable assumptions that the two components have the same abundance and that that the two abundance determinations are independent estimates of the true abundance, then [Fe/H] = -0.075 ± 0.141. For zero age main sequence models, we find that an increase of 0.1 in [Fe/H] gives an increase of 0.0045 in log $R/R_\odot$.

B17 have reconstructed surface brightness distributions from Doppler imaging, and estimate that dark spots cover 1.7% and 5.6% of the surfaces of the A and B components, respectively. In their analysis, they adopt a two-temperature model with the dark regions being 400 K cooler than the bright regions of temperature 2800 K. The equivalent spot coverage fractions for completely dark spots are $f_s$ = 0.8% and 2.6%, respectively. We find that an increase in spot coverage of 1% increases log $R$ for a zero age main sequence model by 3.7 $10^{-4}$. Based on these result, we conclude that the effect of spots on the radii of the GJ 65 components is negligible when compared to the uncertainties from mass and heavy element abundance.

In earlier work (MacDonald & Mullan 2014, 2017a), we have found that predicted surface vertical fields, $B_{surf}^V$, from our GT models scale with the chosen value of $B_{ceil}$ according to $B_{surf}^V \sim B_{ceil}^{-0.07}$. Thus, in GT magneto-convective models (where the magnetic diffusivity is zero), an increase in $B_{ceil}$ by a factor of 100 decreases the surface field required to give the same degree of oversizing by a factor of only ~1.4. However, in our study of the oversized stars in the M7 binary LSPM J1314+1320 (MacDonald & Mullan 2017b), we found that, in GTC magneto-convective models (where the magnetic diffusivity is non-zero), the predicted surface field is independent of the value of $B_{ceil}$. For LSPM J1314+1320, our GTC models required $B_{surf}^V$ = 630 – 1430 G to give the degree of oversizing observed. Because these surface field strengths are much smaller than those observed for GJ 65 A and B, we have calculated new magneto-convective models with higher surface fields and with $B_{ceil}$ values up to 1 MG to determine whether or not the lack of dependence on $B_{ceil}$ for our GTC models still holds. The results of these new models confirm that the GTC surface fields remain independent of the value chosen for $B_{ceil}$.

A long-standing uncertainty in stellar modelling is the choice of mixing length ratio, $\alpha$. For the models used in the following sections, we set $\alpha = 0.7$ for the reasons given in section 3. However, this value of $\alpha$ is much less than that found by solar calibration for which $\alpha \sim 2$. We, therefore, consider the sensitivity of model predictions for stellar radius to the particular choice for $\alpha$. We have shown (MacDonald & Mullan 2010) that the same increase in degree of oversizing and related reduction in $T_{eff}$ can be achieved by either keeping $\alpha$ fixed and increasing $\delta$, or keeping $\delta$ fixed and decreasing $\alpha$. The extent to which the value of $\alpha$ affects the model radii can be seen from figures 7 and 8, which show how the radius at age 5 Gyr depends on $\delta$ for a range of $\alpha$ values. We see that the dependence on $\alpha$ is weaker for larger values of $\delta$.

In the next section, we find, for $\alpha = 0.7$, that the value of $\delta$ needed to match the radius inflation and $T_{eff}$ reduction observed for the components of GJ 65 is ~0.05 and ~0.10 for our GT and GTC models, respectively. From figures 7 and 8, we infer that if the actual value of $\alpha$ was 2.0 then $\delta$ would need to be ~20% and ~12% larger for our GT and GTC models, respectively. The corresponding increase in $B_{surf}^V$ would be ~10% and ~6%, respectively.

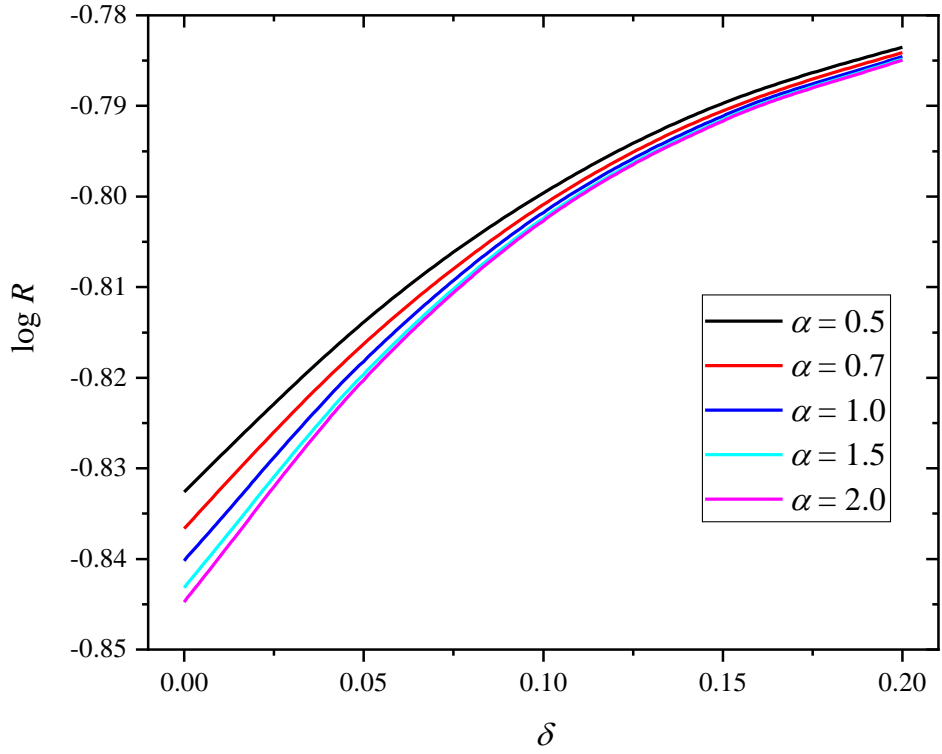

Figure **7**. GT models with various mixing length ratios: Dependence of log $R$ at age 5 Gyr on the magnetic inhibition parameter, $\delta$, for a range of mixing length ratio, $\alpha$ for [Fe/H] = 0.0, $M$ = 0.12 M$_\odot$ models. ($R$ in solar units.)

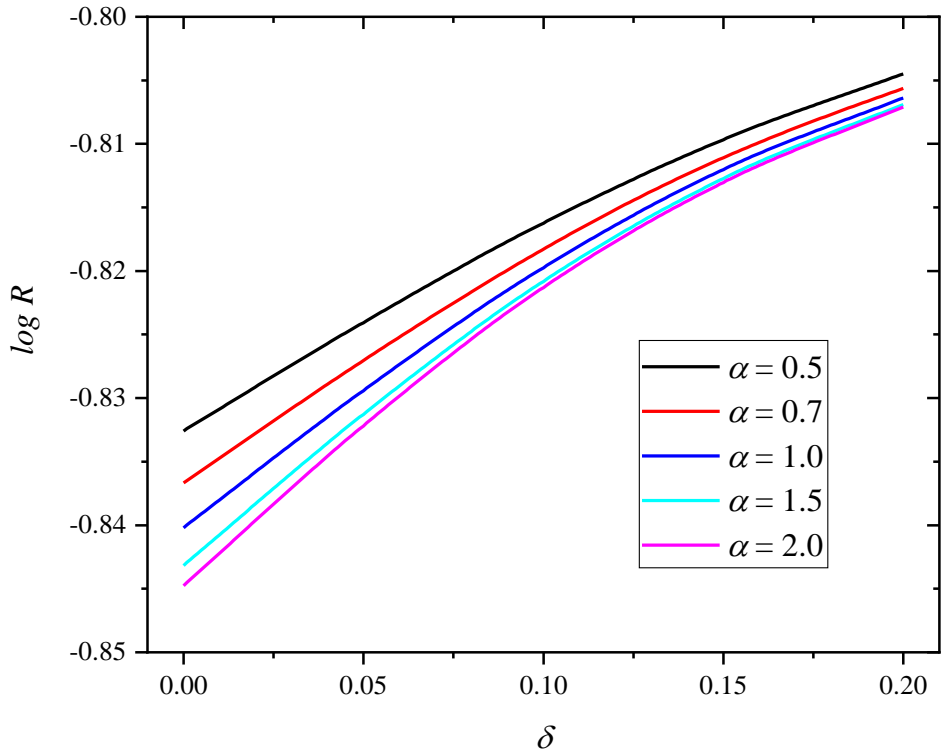

Figure 8. As figure 7, except for GTC models.

## 6. Surface Magnetic Field Strengths

In this section, we derive the strength of the surface magnetic field needed by our models to match for each component of GJ 65 the observed radii and the effective temperatures determined in section 2. We note that for our modelling, the only relevant quantity is the strength of the vertical component of the magnetic field, independent of the scale height or the topology of the field.

The vertical component of the magnetic field, $B_v$, is related to the $\delta$ - parameter by

$$B_v^2 = 4\pi\gamma P_{gas} \frac{\delta}{1-\delta}, \qquad (2)$$

and once a $\delta$ – value has been found by fitting models to the observed radius and/or effective temperature, it is straightforward to find the corresponding photospheric value of $B_v$.

For models of stars of mass in the range $0.11 - 0.15$ M$_\odot$, we find for our 5 Gyr main sequence models that the dependences of log $R/$R$_\odot$, log $T_{eff}$ and log ($\gamma P_{gas}$) on mass, heavy element abundance, spot fraction and magnetic inhibition parameter are well approximated by expressions of form

$$y = a + b_1 \log M + b_2 (\log M)^2 + c[\text{Fe}/\text{H}] + d_1 f_s + d_2 f_s^2 + e_1 \delta + e_2 \delta^2, \qquad (3)$$

where $a$, $b_1$, $b_2$, $c$, $d_1$, and $d_2$ depend only on age, and $e_1$ and $e_2$ depend not only on age but also on whether GT or GTC models are used. In equation (3), $M$ is in units of M$_\odot$. The values of the coefficients for $\alpha = 0.7$ are given in Tables 3 and 4. The units of $P_{gas}$ are dyne cm$^{-2}$.

Table 3. Fit coefficients that are independent of magnetic model

|       | 5 Gyr          |              |                       | 500 Myr        |              |                       |
|-------|----------------|--------------|-----------------------|----------------|--------------|-----------------------|
|       | log $R/$R$_\odot$ | log $T_{eff}$ | log($\gamma P_{gas}$) | log $R/$R$_\odot$ | log $T_{eff}$ | log($\gamma P_{gas}$) |
| $a$   | -0.0556        | 3.1323       | 7.2667                | -0.1183        | 3.2730       | 6.6826                |
| $b_1$ | 0.8485         | -0.9656      | 3.7839                | 0.7736         | -0.6488      | 2.3574                |
| $b_2$ | 0.0            | -0.6480      | 2.8836                | 0.0            | -0.4694      | 2.0095                |
| $c$   | 0.0451         | -0.0673      | -0.656                | 0.0534         | -0.0669      | -0.671                |
| $d_1$ | 0.0456         | -0.102       | 0.318                 | 0.0630         | -0.101       | 0.286                 |
| $d_2$ | 0.0260         | -0.063       | 0.100                 | 0.0610         | -0.0610      | 0.040                 |

Table 4. Fit coefficients that depend on magnetic model

|       | 5 Gyr          |         |              |         |                       |         |
|-------|----------------|---------|--------------|---------|-----------------------|---------|
|       | log $R/$R$_\odot$ |         | log $T_{eff}$ |         | log($\gamma P_{gas}$) |         |
|       | GT             | GTC     | GT           | GTC     | GT                    | GTC     |
| $e_1$ | 0.4598         | 0.2833  | -1.0398      | -0.4706 | 3.0573                | 1.3863  |
| $e_2$ | -1.0268        | -0.1786 | 2.1295       | 0.6207  | -8.2037               | -1.8198 |

|       | 500 Myr        |         |              |         |                       |         |
|-------|----------------|---------|--------------|---------|-----------------------|---------|
|       | log $R/$R$_\odot$ |         | log $T_{eff}$ |         | log($\gamma P_{gas}$) |         |
|       | GT             | GTC     | GT           | GTC     | GT                    | GTC     |
| $e_1$ | 0.5652         | 0.2774  | -0.9861      | -0.4603 | 2.7369                | 1.2469  |
| $e_2$ | -0.2232        | -0.1833 | 1.9716       | 0.5806  | -9.2714               | -1.9966 |

In the reasonable assumption that the effects of dark spots can be ignored (see section 5), given $T_{eff}$, $R$ and [Fe/H], the fits of form in equation (3) can be used to find $M$ and $\delta$.

In table 5 we give the predictions of our models for the stellar mass, surface value of the magnetic inhibition parameter, $\delta$, and the corresponding photospheric vertical magnetic field component, $B^V_{surf}$, when we assume the stars are on the main sequence at age 5 Gyr.

Table 5. Predicted values for stellar mass, magnetic inhibition parameter and surface vertical field at age 5 Gyr.

|  | $T_{eff}$ from K16 radii ||||||  $T_{eff}$ from B17 radii ||||||
|  | GT models ||| GTC models ||| GT models ||| GTC models |||
|  | $M$ (M$_\odot$) | $\delta$ | $B^V_{surf}$ (G) | $M$ (M$_\odot$) | $\delta$ | $B^V_{surf}$ (G) | $M$ (M$_\odot$) | $\delta$ | $B^V_{surf}$ (G) | $M$ (M$_\odot$) | $\delta$ | $B^V_{surf}$ (G) |
|---|---|---|---|---|---|---|---|---|---|---|---|---|
| GJ 65A | 0.1337 ± 0.0042 | 0.045 ± 0.017 | 1134 ± 372 | 0.1307 ± 0.0032 | 0.100 ± 0.042 | 1783 ± 659 | 0.1306 ± 0.0061 | 0.036 ± 0.020 | 993 ± 414 | 0.1282 ± 0.0048 | 0.081 ± 0.049 | 1546 ± 726 |
| GJ 65B | 0.1271 ± 0.0041 | 0.051 ± 0.019 | 1286 ± 400 | 0.1239 ± 0.0031 | 0.116 ± 0.046 | 2047 ± 736 | 0.1278 ± 0.0054 | 0.055 ± 0.022 | 1337 ± 436 | 0.1243 ± 0.0039 | 0.125 ± 0.057 | 2166 ± 851 |

In table 6 we give the predictions of our models for the surface value of the magnetic inhibition parameter, $\delta$, and the corresponding photospheric vertical magnetic field component, $B_v$, when we assume the stars are on the pre-main sequence at age 500 Myr.

Table 6. Predicted values for stellar mass, magnetic inhibition parameter and surface vertical field at age 500 Myr.

|  | $T_{eff}$ from K16 radii ||||||  $T_{eff}$ from B17 radii ||||||
|  | GT models ||| GTC models ||| GT models ||| GTC models |||
|  | $M$ (M$_\odot$) | $\delta$ | $B^V_{surf}$ (G) | $M$ (M$_\odot$) | $\delta$ | $B^V_{surf}$ (G) | $M$ (M$_\odot$) | $\delta$ | $B^V_{surf}$ (G) | $M$ (M$_\odot$) | $\delta$ | $B^V_{surf}$ (G) |
|---|---|---|---|---|---|---|---|---|---|---|---|---|
| GJ 65A | 0.1302 ± 0.0036 | 0.045 ± 0.018 | 1129 ± 372 | 0.1298 ± 0.0035 | 0.101 ± 0.043 | 1752 ± 641 | 0.1275 ± 0.0052 | 0.037 ± 0.021 | 992 ± 413 | 0.1272 ± 0.0051 | 0.081 ± 0.050 | 1523 ± 705 |
| GJ 65B | 0.1229 ± 0.0034 | 0.051 ± 0.019 | 1266 ± 366 | 0.1225 ± 0.0033 | 0.115 ± 0.047 | 1988 ± 702 | 0.1234 ± 0.0044 | 0.057 ± 0.023 | 1319 ± 430 | 0.1230 ± 0.0042 | 0.125 ± 0.057 | 2100 ± 794 |

Comparison of tables 5 and 6 indicates that our predicted field strengths are relatively insensitive to the adopted age. The largest fields are found for our GTC models in which finite diffusivity effects are included. We also find that the predicted fields are larger for GJ 65B than GJ 65A by 10 – 35%, depending on the adopted radii. In regard to field strength ratio our results are consistent with the measurements of Kochukhov & Lavail (2017) and Shulyak et al. (2017), who both find that GJ 65B has a 30% stronger field than GJ 65A. However, the measured field strengths are higher than we predict. We note our models give the strength of the vertical field only. If we assume randomly distributed field directions, then we predict total field strengths of up to $\sim \sqrt{3} B_v$. This leads to total field strengths in our GTC models of up to 3100 ± 1100 G for GJ 65A and 3750 ± 1500 G for GJ 65B, ranges which overlap the field strengths measured by Shulyak et al. (2017).

7. Conclusions and Discussions

By fitting magneto-convective stellar models to the observed radii and inferred effective temperatures of the components of the GJ 65 binary, we have been able to derive values of the field strengths on the surfaces of both components. If the stellar material is assumed to be a perfect conductor, we estimate surface total field strengths of ~1800 ± 700 G for GJ 65A and ~2250 ± 700 G for GJ 65B. When finite conductivity is included, we find that the total field strengths are ~2900 ± 1200 and ~3600 ± 1350 G respectively. These field strengths are significantly larger than the fields we have found in our magneto-convective modelling of other M dwarfs (see Table 1 in MacDonald & Mullan 2017c). The largeness of the fields in GJ65A and B compared to other M dwarfs may be related to their exceptionally rapid rotations: their rotational periods of 0.243 and 0.227 days (B17) make these stars faster rotators than 96% of the 164 M dwarfs in the sample reported by West et al. (2015) and 94% of the 274 M dwarfs in the sample of Newton et al. (2016).

The measured total field strengths in GJ65A and GJ65B are 4.5 ± 1.0 and 5.8 ± 1.0 kG, respectively, according to Shulyak et al. (2017), and 5.2 ± 0.5 and 6.7 ± 0.6 kG respectively according to Kochukhov & Lavail (2017). Although the results obtained by both teams appear to differ, they are actually within 1$\sigma$ of each other. We note that our results for the predicted total field strength on the surfaces of the two stars overlap (within 1$\sigma$) with the results of Shulyak et al. (2017) with one important proviso: the overlap exists only in models where the effects of finite magnetic diffusivity are explicitly incorporated. In models where the diffusivity is assumed to be zero, our surface magnetic field estimates do *not* overlap with the observed field strengths. As far as we are aware, this is the first time that it has been demonstrated that effects of finite conductivity have a detectable effect on the structural properties of a magnetic star.

Our results suggest that finite conductivity cannot be neglected in models of dwarf stars where $T_{eff}$ has values that are as low as 2700 - 2800 K. Is there any reason why magnetic diffusivity should become important at such temperatures? To address this, we note that it becomes necessary to include finite conductivity in magneto-convection if, during the time $t_c$ required for a convective cell (granule) to perform one turn-over, the field lines will diffuse through a distance $L_d$ that is comparable to the size of a granule $L_g$. In a medium where the electrical conductivity is $\sigma_e$ (in electrostatic units), the diffusion length $L_d = c\sqrt{\pi t_c / \sigma_e}$, where $c$ is the speed of light (see Bray & Loughhead 1964). In M dwarfs, $t_c$ is estimated to be ~ 50 s. The value of $L_g$ scales as $T_{eff}/(\mu g)$, where $\mu$ is mean molecular weight and $g$ is gravity (Mullan 1984). In M dwarfs, $T_{eff}$ is less than the solar value by ~2, $\mu$ is a factor of 2 larger and $g$ is larger also, by a factor of ~5. Thus, $L_g$ is expected to be about 20 times smaller than in the Sun (where $L_g$ ~ $10^8$ cm), so that $L_g$ ~ 5 × $10^6$ cm in M dwarfs. The value of $\sigma_e$ that, in the event that $t_c$ = 50 seconds, makes $L_d$ of order 5 × $10^6$ cm is $\sigma_e$ = 6 × $10^9$ esu. Is this a reasonable value to expect in a medium with $T$ = 2700 - 2800 K? To answer this, we use the value of the magnetic diffusivity provided by our stellar evolution code. At the photosphere of a stellar model with $T_{eff}$ = 2750 K, we find $\eta \approx 3.6 \times 10^{10}$ in cgs units, and $\sigma_e$ ~ 3 × $10^9$ esu, and so we conclude that it is plausible to expect that the effects of magnetic diffusivity would make their presence felt in the stars GJ65A and GJ65B. The effects of finite conductivity have also been found to have an effect on the slope of the flare frequency distribution in the coolest flare stars (Mullan & Paudel 2018).

Acknowledgments
This research has made use of the SVO Filter Profile Service (http://svo2.cab.inta-csic.es/theory/fps/) supported from the Spanish MINECO through grant AyA2014-55216, and the VizieR catalogue access tool, CDS, Strasbourg, France. We thank John Gizis for helpful discussions. This work is supported in part by the NASA Delaware Space grant.

References

Allard, F., Homeier, D., & Freytag, B. 2012a, Roy. Soc. London Philos. Trans. Ser. A, 370, 2765
Allard, F., Homeier, D., Freytag, B., & Sharp, C. M. 2012b, in EAS Pub. Ser. 57, eds. C. Reylé, C. Charbonnel, & M. Schultheis, 3
Allard, F., Homeier, D., Freytag, B., Schaffenberger, W., & Rajpurohit, A. S. 2013, MSAIS, 24, 128
van Altena, W. F., Lee, J. T., & Hofleit, E. D. 1995, The general catalogue of trigonometric [stellar] parallaxes)
Baraffe, I., Homeier, D., Allard, F., & Chabrier, G. 2015, A&A, 577, A42
Barnes, J. R., Jeffers, S. V., Haswell, G. A. et al. 2017, MNRAS, 471, 811
Benedict, G. F., Henry, T. J., Franz, O. G., et al. 2016, AJ, 152, 141
Bessell, M. S. 1991, AJ, 101, 662
Boeshaar, P. C., & Davidge, T. J. 1994, The MK process at 50 years. A powerful tool for astrophysical insight, Astronomical Society of the Pacific Conference Series, edited by Chris Corbally, R. O. Gray, and R. F. Garrison, Vol. 60, p246
Bray R. J., & Loughhead, R. E. 1964, Sunspots (New York: Dover), pp. 125, 270
Browning, M. K. 2008, ApJ, 676, 1262
Caffau, E., Ludwig, H.-G., Steffen, M., Freytag, B., & Bonifacio, P. 2011, SoPh, 268, 255
Casagrande, L., Flynn, C., & Bessell, M. 2008, MNRAS, 389, 585
Chuadze, A. D. 1965, Astron. Tsirk. # 345, p. 2-4
Dieterich, S. B., Henry, T. J., Golimowski, D. A., Krist, J. E., & Tanner, A. M. 2012, AJ, 144, 64
Dieterich, S. B., Henry, T. J., Jao, W.-C., et al. 2014, AJ, 147, 94
Dupuy, T. J., & Liu, M. C. 2012, ApJS, 201, 19
Gaidos, E., Mann, A. W., Lépine, S., et al. 2014, MNRAS, 443, 2561
Golimowski, D. A., Leggett, S. K., Marley, M. S., et al. 2004, AJ, 127, 3516
Gough, D. O., & Tayler, R. J. 1966, MNRAS, 133, 85
Henry, T. J., Walkowicz, L. M., Barto, T. C., & Golimowski, D. A. 2002, AJ, 123, 2002
Huber, D., Silva Aguirre, V., Matthews, J. M., et al. 2014, ApJS, 211, 2
Joy, A. H., & Abt, H. A. 1974, ApJS, 28, 1
Karamish, V. F., Migach, Yu. E., & Pogrebnoy, G. D. 1966, Astron. Tsirk. # 392, p. 2-4
Kervella, P., Mérand, A., Ledoux, C., Demory, B.-O., & Le Bouquin, J.-B. 2016, A&A, 593, A127
Kirkpatrick, J. D., Henry, T. J., & McCarthy, D. W., Jr. 1991, ApJS, 77, 417
Kochukhov, O., & Lavail, A. 2017, ApJ, 835, L4
Leger, A., Defrere, D., Malbet, F., Labadie, L., & Absil, O. 2015, ApJ, 808, 194
Leggett, S. K., Allard, F., Berriman, G., Dahn, C. C., & Hauschildt, P. H. 1996, ApJS, 104, 117
Loyd, R. O. P., & France, K. 2014, ApJS, 211, 9
MacDonald, J., & Mullan, D. J. 2010, ApJ, 723, 1599
MacDonald, J., & Mullan, D. J. 2012, MNRAS, 421, 3084
MacDonald, J., & Mullan, D. J. 2014, ApJ, 787, 70
MacDonald, J., & Mullan, D. J. 2017a, ApJ 834, 67
MacDonald, J., & Mullan, D. J. 2017b, ApJ, 843, 142
MacDonald, J., & Mullan, D. J. 2017c, ApJ, 850, 580
Mann, A. W., Gaidos, E., & Ansdell, M. 2013, ApJ, 779, 188
Mann, A. W., Feiden, G. A., Gaidos, E., Boyajian, T., & von Braun, K. 2015, ApJ, 804, 64
Mann, A. W., Feiden, G. A., Gaidos, E., Boyajian, T., & von Braun, K. 2016, ApJ, 819, 87
Migach, Yu. E. 1965, Astron. Tsirk. # 345, p. 1-2
Mullan, D. J. 1984, ApJ, 282, 603
Mullan, D. J., & MacDonald, J. 2010, ApJ, 713, 1249
Mullan, D.J., & Paudel, R. R. 2018, ApJ, 854, 14

Newton, E. R., Irwin, J., Charbonneau, D., et al. 2016, ApJ, 821, 93
Perryman, M. A. C., Brown, A. G. A., Lebreton, Y., et al. 1998, A&A, 331, 81
Rajpurohit, A. S., Reylé, C., Allard, F., et al. 2013, A&A, 556, 15
Reid, I. N., Hawley, S. L., & Gizis, J. E. 1995, AJ, 110, 1838
Schroeder, D.J., Golimowski, D.A., Brukardt, R.A., et al. 2000. AJ, 119, 906
Steffen, M. 1990, A&A, 239, 443
Terrien, R. C., Mahadevan, S., Deshpande, R., & Bender, C. F. 2015, ApJS, 220, 16
Torres, C. A. O., Quast, G. R., da Silva, L., et al. 2006, A&A, 460, 695
West, A. A., Weisenburger, K. L., Irwin, J., et al. 2015, ApJ, 812, 3
Winters, J. G., Henry, T. J., Lurie, J. C., et al. 2015, AJ, 149, 5
Wright, E. L., Eisenhardt, P. R. M., Mainzer, A. K., et al. 2010, AJ, 140, 1868
Wright, N. J., Drake, J. J., Mamajek, E. E., & Henry, G. W. 2011, ApJ, 743, 48
Yadav, R. K., Christensen, U. R., Morin, J., et al. 2015, ApJL, 813, L31